\documentclass[conference]{IEEEtran}
\IEEEoverridecommandlockouts
\def\BibTeX{{\rm B\kern-.05em{\sc i\kern-.025em b}\kern-.08em
    T\kern-.1667em\lower.7ex\hbox{E}\kern-.125emX}}

\usepackage[T1]{fontenc}
\usepackage{graphicx}
\usepackage{hyperref}
\usepackage{color}

\usepackage{amsmath,amssymb,amsfonts}
\usepackage[numbers,sort&compress]{natbib}
\usepackage[most]{tcolorbox}
\usepackage[inline]{enumitem}
\usepackage{booktabs}
\usepackage[table]{xcolor}
\usepackage{subcaption}
\usepackage[para]{footmisc}
\usepackage{tabularx}
\usepackage{pifont}
\usepackage{xspace}
\usepackage{orcidlink}

\newtcolorbox{rqanswer}{
    enhanced,
    colback=gray!20,
}

\newcommand\TOOL{\textsc{CREAD}\xspace}

\newcommand\SIM{\texttt{HighwayEnv}\xspace}

\begin{document}

% \title{Collaborative Red Teaming for Autonomous Driving Safety: A Decentralised Multi-Agent Ecosystem for Emergent Failure Discovery via Perception Fuzzing and Metamorphic Validation}
\title{Collaborative Multi-Agent Testing for Emergent Failure Discovery in Autonomous Driving Systems}

\makeatletter
\newcommand{\linebreakand}{%
  \end{@IEEEauthorhalign}
  \hfill\mbox{}\par
  \mbox{}\hfill\begin{@IEEEauthorhalign}
}
\makeatother

\author{\IEEEauthorblockN{Ruizhen Gu}
\IEEEauthorblockA{
% \textit{School of Electronics, Electrical Engineering and Computer Science} \\
\textit{Queen's University Belfast}\\
Belfast, UK \\
r.gu@qub.ac.uk \\
% \orcidlinkc{0009-0001-8021-7052}
}
\and
\IEEEauthorblockN{Konstantinos Koufos}
\IEEEauthorblockA{
% \textit{dept. name of organization (of Aff.)} \\
\textit{Queen's University Belfast}\\
Belfast, UK \\
k.koufos@qub.ac.uk 
% \\ \orcidlinkc{0000-0001-6616-6212}
}
\and
\IEEEauthorblockN{Donghwan Shin}
\IEEEauthorblockA{
% \textit{School of Computer Science} \\
\textit{The University of Sheffield}\\
Sheffield, UK \\
d.shin@sheffield.ac.uk 
% \\ \orcidlinkc{0000-0002-0840-6449}
}
\linebreakand
\IEEEauthorblockN{Vahid Garousi}
\IEEEauthorblockA{
\textit{Queen's University Belfast}\\
Belfast, UK \\
\textit{Azerbaijan Technical University}\\
Azerbaijan \\
v.garousi@qub.ac.uk
% \\ \orcidlinkc{0000-0001-6590-7576}
}
\and
\IEEEauthorblockN{Mehrdad Dianati}
\IEEEauthorblockA{
% \textit{dept. name of organization (of Aff.)} \\
\textit{Queen's University Belfast}\\
Belfast, UK \\
m.dianati@qub.ac.uk 
% \\ \orcidlinkc{0009-0001-6971-0348}
}
% \and
% %
% \IEEEauthorblockN{6\textsuperscript{rd} Given Name Surname}
}

\maketitle
\thispagestyle{plain}

\begin{abstract}
% Problem 
% Autonomous Driving Systems (ADS) can fail in ways that arise from interactions among perception, planning, and control rather than from faults in a single module. 
Autonomous Driving Systems (ADS) can fail because of faults within individual modules as well as from interactions across perception, planning, and control.
% Gap
% Existing ADS testing methods often improve one function at a time, such as scenario generation, validation, or search, but usually treat these functions as isolated techniques rather than coordinated testing roles.
Yet existing ADS testing research often treats key testing functions, such as perturbation generation, behavioural assessment, and test case selection and exploration, as loosely coupled steps rather than coordinated roles for discovering such failures.
% Contribution
We present \TOOL, a collaborative multi-agent testing framework for testing ADS that organises perturbation generation, behavioural validation, and search coordination through a shared blackboard and an orchestrator. 
% Instantiation
In the current work-in-progress instantiation, the framework focuses on perception-oriented perturbation generation, while remaining extensible to other ADS modules, including planning and control. 
It currently comprises a Perception Fuzzer Agent, a Metamorphic Validator Agent, and an Orchestrator Agent. Respectively, they generate perturbations, assess behavioural consistency across related scenario pairs, and coordinate further exploration. 
% Result
Experiments in \SIM simulator show that the collaborative configuration improves failure discovery in the highway environment and remains competitive in the roundabout setting. 
Across the two environments, it yields about 2.1x as many failures per 100 scenarios as the single-agent baseline on average, while gains over a non-collaborative two-agent baseline vary across environments. 
These results suggest that collaborative multi-agent testing is a promising research direction for emergent ADS behaviour discovery.
\end{abstract}


\section{Introduction}

Autonomous Driving Systems (ADS) are safety-critical, open-world systems whose failures can arise both from faults within individual modules and from interactions among perception, planning, and control~\cite{Lou2022TestingADS, machines5010006}. Given this complexity, verification and validation for ADS increasingly relies on scenario-based testing to build safety-relevant evidence~\cite{riedmaier2020}. In this paradigm, functional scenarios derived from the operational design domain (ODD) are progressively refined into concrete scenarios, which can then be executed as test cases in simulation or other test environments~\cite{Menzel2018}. Within such a pipeline, effective testing requires not only meaningful perturbation generation, but also reliable behavioural assessment and efficient exploration of the scenario space. Yet recent testing approaches based on feedback-guided fuzzing, adaptive search, and LLM-assisted scenario synthesis still struggle to consistently uncover rare but high-consequence failures, especially those arising in the long tail of safety-critical driving scenarios~\cite{AVFuzzer2020,DriveFuzz2022,LeGEND2024}. 
While some failures, such as collisions, are easy to recognise, more subtle behavioural degradations are harder to assess consistently, especially when they emerge from interactions across multiple modules~\cite{Molina2025Oracle}.

This work is motivated by two limitations of current ADS testing. 
First, recent approaches of scenario generation, such as LLM-guided synthesis, improve realism and efficiency, but still struggle to steer exploration towards diverse and safety-relevant scenario families without repeatedly generating similar test cases~\cite{Lou2022TestingADS,DriveFuzz2022,LeGEND2024}. 
Second, current ADS testing pipelines often treat scenario generation and behavioural validation as separate or only loosely coupled stages, limiting feedback between the creation of candidate scenarios and the assessment of their safety relevance~\cite{GenAIADSSurvey2026}.
We address these limitations by treating ADS testing as a collaborative process rather than as a single generation-and-evaluation loop.

Drawing on agent-based software testing~\cite{10440574}, we frame scenario-based safety-focused testing for ADS as a collaborative testing problem and introduce \TOOL (\textbf{C}ollabo\textbf{R}ative for \textbf{E}mergent \textbf{A}DS behaviour \textbf{D}iscovery). 
Unlike prior scenario-generation frameworks, including multi-agent and LLM-assisted approaches, which mainly emphasise realistic and efficient construction of challenging scenarios~\cite{AMACollision2025,LeGEND2024}, \TOOL focuses on coordinating perturbation generation, validation, and exploration to improve failure yield. 
% It adopts a closed-loop process in which perturbation generation, behavioural validation, and adaptive coordination operate as interacting testing roles. 
In the present instantiation, this process is realised through three coordinated roles: 
\begin{enumerate*}[label=(\arabic*)]
    \item Perception Fuzzing Agent for generating perception-oriented perturbations and scenario variants;
    \item Metamorphic Validation Agent for comparing baseline and perturbed executions to detect safety-relevant behavioural inconsistencies; and 
    \item Orchestrator Agent that prioritises promising scenario families for further testing through a decentralised blackboard architecture.
\end{enumerate*}
% The Perception Fuzzing Agent generates perception-oriented perturbations and scenario mutations to propagate through the ADS pipeline and induce downstream safety-relevant deviations. The Metamorphic Validation Agent evaluates whether related scenario variants produce behavioural changes that are disproportionate to the intended semantic differences, helping to identify potentially unsafe or unstable system responses~\cite{DriveFuzz2022,DeepRoad2018,Deng2023DeclarativeMT}. 
% Finally, the Orchestrator Agent maintains a Quality-Diversity archive inspired by MAP-Elites~\cite{mouret2015illuminatingsearchspacesmapping} and directs testing effort towards underexplored yet promising scenario families.
These roles interact through a decentralised blackboard architecture, forming a closed-loop testing process in which agents continually inform one another. 

A key advantage of this collaborative design is that it allows complementary testing functions to inform one another during search rather than operating as isolated steps. For example, the Perception Fuzzer may generate a mild-glare motorway-merge scenario, and the Metamorphic Validator may then detect delayed braking or reduced time-to-collision relative to the baseline. This feedback is written to the blackboard and used by the Orchestrator to prioritise similar or nearby scenarios in subsequent explorations.

In our evaluation using \SIM, in the Highway environment, the collaborative configuration increases the number of collision-causing scenarios from 14 to 52 per 100 tested scenarios relative to a non-collaborative multi-agent baseline, while matching failure-type diversity and increasing the perception fault rate.
In the Roundabout environment, it remains competitive, yielding 52 failures per 100 scenarios compared with 58 for the non-collaborative baseline, while slightly improving failure diversity and perception fault rate. Relative to a single-agent baseline, the gains in failure discovery are also substantial, particularly in Highway.

In this work, perception fuzzing serves as a proof-of-concept instantiation rather than a restriction of the framework. The architecture is designed to support other testing roles, including agents targeting planning and control.
The current implementation is evaluated in a controlled simulation, with comprehensive evaluation left to future work.
The implementation is available at: \url{https://github.com/ruizhengu/CREAD}

This work makes two contributions:
\begin{itemize}
    \item It introduces \TOOL as a collaborative multi-agent testing paradigm for ADS, coordinating distinct testing roles through a shared blackboard and an adaptive orchestrator rather than relying on a single end-to-end search engine.
    \item It provides empirical evidence that the collaborative testing improves failure discovery over simpler baselines. 
    In our study, the collaborative configuration increases failure discovery from 12 to 52 failures per 100 scenarios in the highway and remains competitive in the roundabout setting, outperforming the single-agent baseline while also revealing environment-dependent tradeoffs against a non-collaborative multi-agent baseline.
\end{itemize}

\section{Background and Related Work}

\subsection{ADS Fuzzing}

Coverage-oriented and adversarial fuzzing methods have significantly advanced ADS testing by automating the discovery of safety-critical scenarios in large input spaces. 
AV-Fuzzer introduced search-based violation discovery in simulated driving and showed that guided mutation can effectively expose safety violations~\cite{AVFuzzer2020}. 
DriveFuzz further improved this line of work by incorporating driving-quality feedback to guide fuzzing towards vulnerable trajectories more efficiently~\cite{DriveFuzz2022}. 
More recent studies, including LLM-guided scenario generation, highlight the potential of language-driven synthesis for constructing diverse and realistic edge cases~\cite{LeGEND2024}.

However, these approaches mainly strengthen scenario generation itself and are typically organised as a single generation-and-evaluation loop. This leaves limited support for coordinating multiple testing roles or reusing intermediate evidence across strategies. Our work addresses this gap by treating fuzzing as one component within a collaborative testing architecture for failure discovery.

\subsection{Metamorphic Testing}

The oracle problem remains a major bottleneck in ADS validation, especially when complete ground-truth labels are unavailable~\cite{ADSOracleJahangirova}.
Metamorphic testing addresses this by specifying expected consistency relations between transformed but semantically equivalent inputs. 
DeepRoad is an early representative example that uses environmental transformations to expose inconsistent driving behaviour~\cite{DeepRoad2018}. 
More recent work extends this direction through declarative metamorphic frameworks for autonomous driving~\cite{Deng2023DeclarativeMT}.

These methods are effective for identifying behaviour-level inconsistencies, but they are often used as standalone validation mechanisms. As a result, metamorphic signals usually play a limited role in guiding subsequent test generation. Our work addresses this gap by treating metamorphic validation as an interacting testing role within the overall discovery loop.

\subsection{Multi-Agent ADS Testing}

Collaborative and multi-agent approaches have recently gained traction in ADS testing, including multi-agent reinforcement-learning methods for generating adversarial traffic interactions~\cite{AMACollision2025}. These approaches can improve the realism and efficiency of critical scenario generation. More broadly, prior surveys argue that ADS assurance requires combining complementary testing paradigms rather than relying on isolated techniques~\cite{Lou2022TestingADS,ding2023survey}. 

Existing multi-agent approaches, however, still focus mainly on generation. They provide limited support for collaboration among complementary testing roles such as perturbation generation, behavioural validation, and adaptive coordination. Our work addresses this gap through a blackboard-based architecture in which specialised agents exchange partial evidence and an orchestrator reallocates effort during failure discovery.
\section{Approach} \label{sec:approach}

\begin{figure}
    \centering
    \includegraphics[width=1\linewidth]{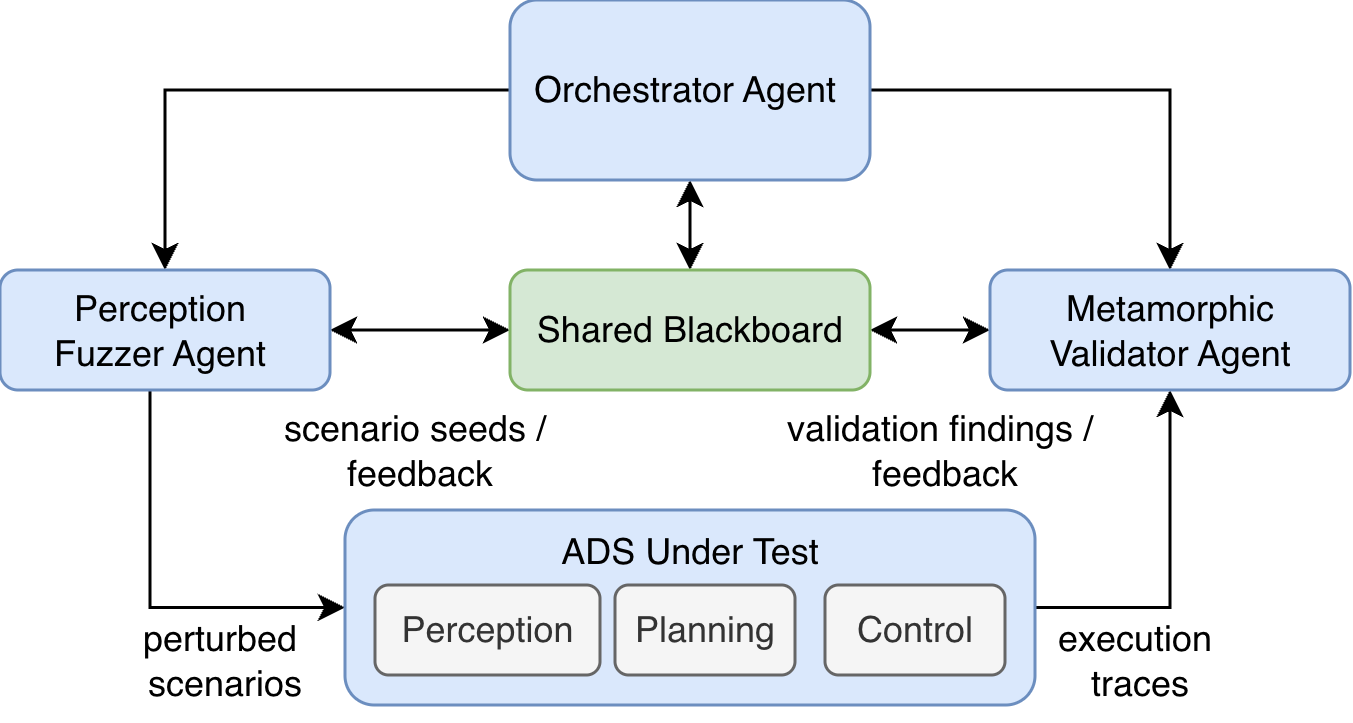}
    \caption{Overview of \TOOL.}
    \label{fig:overview}
\end{figure}

This section presents \TOOL, a collaborative multi-agent framework for ADS testing. The framework organises scenario perturbation, behavioural validation, and adaptive coordination as interacting testing roles in a closed loop. Figure~\ref{fig:overview} shows the overall architecture and the interactions among the Orchestrator, the shared blackboard, and the testing agents.

At each iteration, the Orchestrator selects or prioritises a baseline scenario and allocates testing effort. The Perception Fuzzer Agent proposes targeted perturbations intended to induce safety-relevant deviations. The simulator executes the baseline and perturbed scenarios, and the Metamorphic Validator Agent compares the resulting behaviours to determine whether the perturbation causes disproportionate behavioural changes. The resulting evidence is written to the blackboard and reused by the Orchestrator to guide subsequent testing.

\subsection{Blackboard Collaboration Mechanism}

The blackboard is the framework's shared coordination mechanism. It stores baseline scenarios, perturbed scenarios, execution traces, validation outcomes, and summary coverage information, allowing agents to exchange intermediate evidence asynchronously without direct point-to-point coupling. This shared-state design keeps the agents loosely coupled and makes the framework extensible with additional testing roles.

\subsection{Orchestrator Agent}

The Orchestrator Agent coordinates global testing priorities through a coarse QD-inspired archive over scenario and outcome descriptions. 
Rather than generating perturbations or performing validation itself, it selects which baseline scenarios to revisit or mutate based on heuristic scores derived from prior execution feedback. 
In the current implementation, these scores combine coarse indicators such as failure yield, severity, novelty, and diversity. Guided by these signals, the orchestrator prioritises seeds from underexplored but promising archive regions and allocates further testing effort to scenario families that appear likely to expose additional failures. % This process is archive-guided and heuristic rather than formally optimised.

\subsection{Perception Fuzzer Agent}

The Perception Fuzzer Agent generates structured test scenarios intended to stress the ADS under test. In the current implementation, it starts from a seeded scenario template. It applies LLM-guided refinement to produce a scenario specification that includes the traffic configuration, the ego vehicle's initial state, adversarial vehicles, and a perception configuration. 
The perception configuration includes presets such as fog, rain, night, or dusk, which are then mapped to the perturbation model used at execution time.

Each generated scenario is evaluated in two passes. The framework first executes the scenario as a baseline run without perception perturbation, and then executes the same scenario again with the perturbation specified by its perception configuration.
This design enables comparisons of behavioural differences under matched scenario conditions, rather than across independently generated scenes.

The agent records each generated scenario, its perturbation context, and the resulting outcomes, shared through the blackboard to support metamorphic validation and later scenario selection by the orchestrator. 
In this way, the agent functions as one testing role within a collaborative loop, rather than as a standalone scenario generator.

\subsection{Metamorphic Validator Agent} \label{sec:approach:metamorphic}

The Metamorphic Validator Agent provides behavioural assessments for the framework. It compares baseline and perturbed executions to determine whether a scenario variation leads to disproportionate changes in safety-relevant behaviour. 
In the current implementation, the comparison is based on execution-level summaries such as collision, near-miss occurrence, and perception-related faults, where the latter denotes safety-relevant discrepancies between the simulator ground truth and the perturbed perceived state induced by the Fuzzer.
A violation is reported when the perturbed run exceeds rule-based thresholds relative to baseline, for example, through increased collision rate, run under the same scenario structure. 
The resulting signal is therefore an indicator of behavioural inconsistency rather than a complete formal oracle.

Based on this framework, we study two research questions in the evaluation: 
\begin{itemize}
    \item \textbf{RQ1:} Does agent collaboration improve failure discovery compared to non-collaborative configurations with the same number of agents?
    \item  \textbf{RQ2:} Does a non-collaborative multi-agent configuration improve failure discovery over a single-agent baseline?
\end{itemize}
\section{Evaluation}

This section evaluates whether the proposed collaborative testing architecture improves failure discovery in ADS testing. Our evaluation is designed to address the research questions proposed in Section~\ref{sec:approach}. To answer these questions, we conduct an ablation study under controlled simulation conditions.

\subsection{Evaluation Methodology}

We conduct an ablation study to isolate the contribution of collaborative and multi-agent architecture. Three configurations are evaluated:
\begin{enumerate*}[label=(\arabic*)]
    \item \textbf{Collaborative (Full System):} Both testing agents (Perception Fuzzer and Metamorphic Validator) with blackboard and orchestration;
    \item \textbf{Non-collaborative:} Both testing agents without blackboard or orchestration, executed in sequential order;
    \item \textbf{Single-Agent (Baseline):} Perception Fuzzer agent only, without collaboration infrastructure.
\end{enumerate*}

\begin{table}
    \rowcolors{1}{white}{gray!20}
    \centering
    \begin{tabular}{lccc} \toprule
        \textbf{Configuration} & \textbf{Agents} & \textbf{Blackboard} & \textbf{Orchestrator} \\ \midrule
        Collaborative (Full System) & F + M & \checkmark & \checkmark \\
        Non-collaborative & F + M & \text{\sffamily X} & \text{\sffamily X} \\
        Single-Agent (Baseline) & F & \text{\sffamily X} & \text{\sffamily X} \\ \bottomrule
    \end{tabular}
    \caption{Study configurations (F = Perception Fuzzer Agent, M = Metamorphic Validator Agent). Both Collaborative and Non-collaborative configurations employ the same agents, differing in whether collaboration infrastructure is enabled.}
    \label{tab:configurations}
\end{table}

Table~\ref{tab:configurations} summarises these configurations. The comparison between Collaborative and Non-collaborative isolates the effect of inter-agent collaboration, while the comparison between Non-collaborative and Single-Agent isolates the contribution of adding multiple testing agents without collaboration.

Each configuration is executed for 50 iterations in each environment. The study is intended to evaluate the testing architecture as a failure-discovery mechanism rather than to benchmark driving quality or controller performance.

\subsection{Experimental Setup}

\subsubsection{Simulator and ADS Under Test}

\SIM~\cite{highway-env} is used as the controlled ablation platform. It provides several compact driving scenarios, such as \emph{Highway}, \emph{Roundabout}, \emph{Merge}, and \emph{Intersection}. 
We select \emph{Highway} and \emph{Roundabout} because they offer two complementary interaction profiles: a structured straight-road environment and a denser roundabout with richer vehicle interactions.

We choose \SIM because it supports efficient repeated execution and direct access to traffic state under reproducible conditions, making it well-suited for comparing testing architectures in a controlled setting. Surrounding traffic follows the simulator’s built-in microscopic behaviour models, including IDM-style (intelligent driver model) dynamics~\cite{Kesting_2010}.

To isolate the effect of the testing architecture, we keep the ADS under test simple and fixed across all ablations. 
The ego vehicle is controlled by a deterministic heuristic policy supported by \SIM\footnote{\url{https://highway-env.farama.org/dynamics/vehicle/behavior/}}, selecting lane-change and speed-control actions at each simulation step.
For evaluation, each generated scenario is executed twice, once without perturbation and once with the derived perturbation, with each rollout capped at 50 simulation steps.
The ego policy should therefore be viewed as a rule-based behaviour-selection baseline rather than a learned controller~\cite{Kesting2007MOBILG, SHARMA2021104211}.

\subsubsection{Perception Perturbation Model}

\SIM does not provide an explicit perception stack. To study perception-related failures, we model perception degradation as structured corruption of the ego vehicle's perceived state derived from simulator ground truth, including both the ego vehicle's own perceived state and safety-critical cues about nearby traffic.
The perturbation module injects safety-relevant distortions such as lane-offset and speed-estimation noise, and position error. These perturbations are then consumed by the ego policy and by the fault-analysis metrics, enabling us to study how perception errors propagate into planning and control behaviour under controlled conditions.

This setup should be interpreted as a state-estimation-level proxy for perception degradation rather than a photorealistic sensor simulation. Its purpose is to support architecture-level ablation of collaborative testing behaviour, not to replace high-fidelity validation in a simulator such as CARLA~\cite{Dosovitskiy17}.

\subsubsection{AI Models}

In the current evaluation, the Perception Fuzzer Agent is the only component that uses an LLM. It uses z.ai's \texttt{GLM-5-Turbo}\footnote{\url{https://docs.z.ai/guides/llm/glm-5-turbo}} through API calls to refine seeded scenarios into targeted perception-oriented test cases. 
Although the broader framework can accommodate LLM-based reasoning in other agents, those capabilities are not part of the reported ablation setup.

\begin{table*}[t]
    \centering
    \rowcolors{2}{white}{gray!20}
    \begin{tabular}{lcccccc} \toprule
        & \multicolumn{3}{c}{\textbf{Highway}} & \multicolumn{3}{c}{\textbf{Roundabout}} \\
        \cmidrule(lr){2-4} \cmidrule(lr){5-7}
        \textbf{Metric} & \textbf{Collaborative} & \textbf{Non-collaborative} & \textbf{Single-Agent} & \textbf{Collaborative} & \textbf{Non-collaborative} & \textbf{Single-Agent} \\ \midrule
        Failures/100 scenarios & \textbf{52.00} & 14.00 & 12.00 & 52.00 & \textbf{58.00} & 38.00 \\
        Unique failure types & \textbf{3.9} & \textbf{3.9} & 3.7 & \textbf{3.8} & 3.7 & \textbf{3.8} \\
        Collision rate & \textbf{12.98\%} & 1.98\% & 0.78\% & \textbf{15.66\%} & 11.45\% & 6.78\% \\
        Near-miss rate & \textbf{44.22\%} & 28.79\% & 27.49\% & 21.16\% & \textbf{23.92\%} & 20.03\% \\
        Perception fault rate & \textbf{20.07\%} & 14.30\% & 15.25\% & 19.08\% & 15.77\% & \textbf{19.46\%} \\ \bottomrule
    \end{tabular}
    \caption{LLM-enabled ablation results across the Highway and Roundabout environments (50 iterations each).}
    \label{tab:results_combined}
\end{table*}

\subsection{Evaluation Metrics}

As the goal of the framework is failure discovery rather than driving performance optimisation, our primary metric is \emph{failures per 100 scenarios}, which measures how often a testing configuration exposes failure-causing scenarios.
To better characterise the discovered failures, we also report \emph{unique failure types}, which count how many distinct predefined failure categories are triggered during execution. In the current study, these categories are collision, near-miss, lane departure, speed violation, and perception fault (discussed in Section~\ref{sec:approach:metamorphic}). We further report \emph{perception fault rate}, which reflects sensitivity to simulated perception degradation.
Finally, we include \emph{collision rate} and \emph{near-miss rate} as supporting indicators of the severity profile of the discovered scenarios. 

\subsection{RQ1: Impact of Inter-Agent Collaboration}

Table~\ref{tab:results_combined} summarises the final LLM-enabled ablation results for Highway and Roundabout over 50 iterations per configuration. 
To assess the effect of inter-agent collaboration, we compare Collaborative and Non-collaborative multi-agent settings, which differ in whether they use blackboard-based coordination and orchestration.

The results show that the effect of collaboration is positive, but it depends on the environment. In Highway, the collaborative configuration substantially improves failure discovery, increasing failures from 14 to 52 per 100 scenarios. It also matches the non-collaborative baseline on unique failure types (3.9 vs.\ 3.9).
In Roundabout, the collaborative configuration does not improve failure discovery, yielding 52 failures per 100 scenarios compared with 58 for the non-collaborative baseline. However, it slightly improves failure diversity (3.8 vs.\ 3.7).

Overall, these results suggest that blackboard-based coordination can strengthen failure discovery in some environments, but its benefit is not consistent across all settings.

\begin{rqanswer}
\textbf{Answer to RQ1:} Agent collaboration improves failure discovery in Highway, but not in Roundabout. In Highway, it increases failures from 14 to 52 per 100 scenarios while maintaining the same failure diversity and raising the perception fault rate. In Roundabout, it does not outperform the non-collaborative baseline in failure frequency, although it slightly improves failure diversity and perception fault rate.
\end{rqanswer}

\subsection{RQ2: Contribution of Multi-Agent Architecture}

We compare the Non-collaborative and Single-Agent configurations to evaluate whether adding multiple testing agents, without collaboration, improves failure discovery over a single-agent baseline.

The results show that the non-collaborative multi-agent configuration improves failure discovery in both environments, although the size of the gain varies. In Highway, failures increase from 12 to 14 per 100 scenarios, and failure diversity also rises slightly from 3.7 to 3.9. However, the perception fault rate decreases slightly from 15.25\% to 14.30\%. In Roundabout, the improvement in failure discovery is much larger, with failures increasing from 38 to 58 per 100 scenarios. The two configurations remain similar in failure diversity (3.7 vs. 3.8), while the perception fault rate is lower in the non-collaborative multi-agent setting (15.77\% vs. 19.46\%).

Overall, these results suggest that adding multiple agents can improve failure discovery even without collaboration, especially in the Roundabout environment, although the gains are less consistent across the other metrics.

\begin{rqanswer}
\textbf{Answer to RQ2:} The non-collaborative multi-agent configuration improves failure discovery over the single-agent baseline in both environments. In Highway, failures increase from 12 to 14 per 100 scenarios, while in Roundabout, they increase more substantially from 38 to 58. However, improvements in failure diversity and perception fault rate are less consistent.
\end{rqanswer}

\subsection{Discussion}

The evaluation shows that the proposed architecture is promising as a failure-discovery framework, but that its advantages depend on the traffic environment and on the comparison baseline. 
Inter-agent collaboration shows benefit in Highway, which substantially increases failure yield while maintaining failure diversity and increasing perception fault rate. 
In Roundabout, collaboration remains competitive, but does not dominate the non-collaborative multi-agent configuration on the primary failure-discovery metric.

A second observation is that multiple testing roles are beneficial even without collaboration. The comparison between Non-collaborative and Single-Agent shows that adding a testing role can improve failure discovery, particularly in Roundabout. This suggests that the architectural decomposition of testing functions already contributes value, while collaboration further shapes how effectively those roles interact.

Overall, the results support collaborative testing as a useful architectural direction for ADS failure discovery, while also indicating that interaction-heavy environments remain more sensitive to coordination strategy and search design.

\subsection{Limitations}

\textbf{Perception Abstraction:} \SIM does not provide an explicit full sensor simulation, such as a camera or LiDAR. To study perception-related failures, we approximate perception degradation by injecting structured perturbations into the ego vehicle's derived perceived state. This allows controlled analysis of how perception errors propagate into downstream behaviour, but does not reflect a full-stack perception.

\textbf{Controlled Ablation Platform:} 
\SIM is a lightweight traffic simulator and does not provide the realism of higher-fidelity ADS platforms such as CARLA~\cite{Dosovitskiy17}. We use it as a controlled ablation environment to enable fast, repeatable evaluation of the testing architecture, but the external validity of the results remains limited. Extending the framework to higher-fidelity simulators such as CARLA is part of our future work.

\textbf{Heuristic Ego Policy:} 
The ego vehicle is controlled by a deterministic heuristic policy rather than a production ADS stack or a learned end-to-end controller. We use this simplified policy to isolate the effect of the testing architecture, but the resulting failures should be interpreted as evidence of architectural behaviour under controlled conditions rather than as realistic deployed ADS failures.

\textbf{ODD-Constrained Perturbations:}
In the current version of \textsc{CREAD}, perturbations are designed to remain within the operational design domain (ODD) of the underlying driving setting. This allows us to study whether safety-relevant failures can emerge under plausible in-domain variation, rather than from explicitly out-of-domain conditions. As a result, the current evaluation does not yet assess how the framework behaves when scenarios cross ODD boundaries or expose the ADS to clearly unsupported conditions.
\section{Conclusion and Future Work}

This paper presented \TOOL, a collaborative multi-agent testing framework for Autonomous Driving Systems. It organises perturbation generation, behavioural validation, and coordination as interacting testing roles, rather than treating ADS testing as a single generation-and-evaluation loop. Our evaluation in \SIM shows that the benefits of collaboration are positive but environment-dependent.
In the Highway environment, the collaborative configuration improves failure discovery over the non-collaborative multi-agent baseline, increasing failures from 14 to 52 per 100 scenarios. 
In Roundabout, it remains competitive but does not outperform the non-collaborative baseline. 
We also find that the non-collaborative multi-agent configuration improves failure discovery over the single-agent baseline in both environments.
Overall, these results suggest that collaborative testing is a promising direction for ADS failure discovery.

Future work will extend the framework in three directions. First, we will evaluate it in higher-fidelity simulators such as CARLA to assess its effectiveness under more realistic sensing and traffic conditions. Second, we will incorporate more advanced ADS controllers, including learning-based driving agents, to study the framework beyond the current heuristic ego policy. Third, we will expand the evaluation to additional driving scenarios and conduct a more comprehensive empirical study of the framework's strengths and limitations.

\section*{Acknowledgement}

This work constitutes a part of the HIVEMIND project funded by the European Commission under the Horizon Europe call HORIZON-CL4-2024-DIGITAL-EMERGING-01 under Grant Agreement Number 101189745.

% \newpage

\bibliographystyle{IEEEtranN}
\bibliography{reference}

\end{document}